\documentclass[a4paper,11pt]{article}
\usepackage{jheppub} 
\usepackage{lineno}
\usepackage[english]{babel}
\usepackage{simplewick}
\usepackage{verbatim}
\usepackage{epsfig}
\usepackage{latexsym}
\usepackage{amsmath}
\usepackage{amssymb}
\usepackage{amsfonts}
\usepackage{bm,bbm}
\usepackage{graphicx}
\usepackage{appendix}
\usepackage[dvipsnames]{xcolor}
\usepackage{multirow}
\usepackage{blindtext, rotating}
\usepackage[pdftex]{pict2e}
\usepackage{slashed}
\usepackage{physics}
\usepackage{leftidx}

\usepackage{hyperref}
\usepackage{cleveref}
\usepackage{comment}

\newcommand{\be}{\begin{equation}}
\newcommand{\ee}{\end{equation}}
\newcommand{\bea}{\begin{eqnarray}}
\newcommand{\eea}{\end{eqnarray}}
\newcommand{\id}{\mathbbm{1}}

\def\bsxi{{\boldsymbol \xi}}

\newcommand{\lb}{\left\lbrace}
\newcommand{\rb}{\right\rbrace}
\newcommand*\diff{\mathop{}\!\mathrm{d}}
\newcommand{\rmi}[1]{{\mbox{\scriptsize #1}}}
\newcommand{\rmii}[1]{{\mbox{\tiny\rm{#1}}}}

\title{Baryonic thermal screening mass at NLO}

\author[a,b]{Leonardo Giusti,}
\emailAdd{leonardo.giusti@unimib.it}

\author[c]{M.~Laine,}
\emailAdd{laine@itp.unibe.ch}

\author[a,b]{Davide Laudicina,}
\emailAdd{d.laudicina1@campus.unimib.it}

\author[b]{Michele Pepe,}
\emailAdd{michele.pepe@mib.infn.it}

\author[a,b]{Pietro Rescigno}
\emailAdd{p.rescigno1@campus.unimib.it}

\affiliation[a]{Department of Physics
``Giuseppe Occhialini'', University of Milano-Bicocca,\\ 
Piazza della Scienza 3, I-20126 Milano, Italy}

\affiliation[b]{INFN Milano--Bicocca,\\ 
Piazza della Scienza 3, I-20126 Milano, Italy}

\affiliation[c]{AEC, 
Institute for Theoretical Physics, 
University of Bern, \\ 
Sidlerstrasse 5, CH-3012 Bern, Switzerland}

\abstract{We determine the resummed 1-loop correction 
to a baryonic thermal screening mass. The calculation is 
carried out in the framework of a dimensionally reduced effective theory, 
where quarks are heavy fields due to their non-zero Matsubara
frequencies. The correction due to interactions 
is computed at O($g^2_{ }$) in the coupling constant. In order to solve 
a 3-body Schr\"odinger equation, we exploit a two-dimensional generalization 
of the hyperspherical harmonics method. At electroweak scale
temperatures, the NLO correction represents
a $\sim 4.6 \%$ increase of the 
free-theory value $3\pi T$ of the screening mass.}

\begin{document}
\maketitle
\flushbottom

%
\section{Introduction}

The dynamics of QCD at high temperatures plays
a r\^ole in a number of physical processes,  
ranging from the cosmological evolution of the universe 
a few microseconds after the Big Bang, to the interpretation
of empirical data from heavy ion collision experiments. 
Due to asymptotic freedom, one may hope that at high enough temperatures, 
the influence of interactions can be incorporated 
via a weak-coupling expansion. However, 
QCD effectively behaves as a three-dimensional
Yang-Mills theory in this regime~\cite{Linde:1980ts}, 
which displays confinement. 
This implies that the coefficients appearing in the weak-coupling
expansion need to be solved non-perturbatively, starting from 
some order. Sometimes, it is possible to go to a high order
before non-perturbative coefficients appear, for instance 
$O(g^6_{ }T^4_{ }\ln(1/g))$ for the equation of 
state~\cite{Braaten:1995jr,Kajantie:2002wa}. 
In other cases, such as the Debye screening mass associated with
certain gluonic operators, only the level
$O(g^2_{ }T\ln(1/g))$ can be reached perturbatively, 
before non-perturbative effects 
are met~\cite{Rebhan:1993az,Arnold:1995bh}.

Apart from the appearance of non-perturbative coefficients, 
another issue with thermal perturbation theory is 
that the convergence of the weak-coupling expansion appears to be 
slow, up to very high temperatures. 

In the present study, we focus on screening masses
in the {\em hadronic sector}. Screening masses are the inverses of spatial
correlation lengths, which describe how the quark-gluon plasma reacts
when a state with given quantum numbers is put into the system. 
For hadronic observables, screening is very efficient, with a large
mass of $O(\pi T)$ appearing at leading order. The next-to-leading
order (NLO) 
correction is of $O(g^2_{ }T)$, without any logarithm. 
This correction turns out to be perturbative, i.e.\ infrared (IR) finite, 
even though individual diagrams do contain IR divergences.

As usual, hadronic operators can be divided into mesonic and
baryonic ones. The complete $O(g^2_{ }T)$ correction to flavour 
non-singlet mesonic screening masses was determined a long time ago, 
in the framework of a dimensionally reduced effective 
theory~\cite{Laine:2003bd}. Recently, lattice calculations of the mesonic
masses have reached an unprecedented accuracy over a wide range of
temperatures, and allowed for a direct comparison between perturbative 
and non-perturbative 
data~\cite{Bazavov:2019www,DallaBrida:2021ddx}.  
Lattice data were
found to be compatible with the $O(g^2_{ }T)$ correction, even if 
fast convergence seems to be limited to temperatures 
which are well above the electroweak scale. 

On the other hand, the baryonic sector has been investigated 
with less precision, 
both on the analytical and on the numerical side. 
The analytical result available in the literature is qualitative
\cite{Hansson:1994nb}, and all the lattice calculations, both in the
quenched approximation \cite{Detar:1987hib,Gocksch:1987nt} and in the
full theory \cite{Gottlieb:1987gz,Gupta:2013vha}, are restricted to
very low temperatures, at which a direct comparison with 
analytical results seems difficult. Moreover, no continuum-limit
extrapolation has been performed. Nonetheless, the steady
progress in lattice measurements at very high temperatures
\cite{DallaBrida:2021ddx,Rescigno:2023tkq,Giusti:2024ohu} 
motivates a quantitative
estimate of the NLO correction to baryonic
screening masses, 
and this is the main goal of this paper.

Our presentation is organized as follows.  
In sec.~\ref{sec:preliminaries}, we define the correlation functions 
we are interested in, and give the definition of screening masses,
which characterize their asymptotic behaviour at large distances. 
Section~\ref{sec:eft} introduces the dimensionally 
reduced effective theory of QCD, as an efficient tool for resummed
perturbative computations. In sec.~\ref{sec:bar},  
the correlation functions from sec.~\ref{sec:preliminaries} 
are re-expressed in the language of the effective theory, 
and they are shown to satisfy 
a two-dimensional Schr\"odinger equation. 
A numerical solution of the Schr\"odinger equation is 
presented in sec.~\ref{sec:schrodinger}, before we turn to 
conclusions in sec.~\ref{sec:conclusions}.
Further details on the effective theory, on the extraction
of a static potential, and on the
methods used for the numerical solution of the
Schr\"odinger equation, are reported in four appendices. 
For QCD, we adopt the notation reported in appendix~B of
ref.~\cite{DallaBrida:2020gux} and, unless stated otherwise, 
refer to it for unexplained notation.

%
\section{Preliminaries}
\label{sec:preliminaries}

We are interested in an interpolating operator which carries the
nucleon quantum numbers. Maybe the simplest is
(we consider $N^{ }_{\rm c} = 3$ colours throughout)
\begin{align}
 \label{eq:nucleon}
 N_\alpha^{ }  = 
 \epsilon^{abc}_{ }
 \bigl( u^{a\rm{T}}_{ } C\gamma_5^{ } d^{b}_{ } \bigr)
 d_\alpha^c
 \,,
 \qquad
 \overline{\!N}_\alpha^{ }  = 
 \epsilon^{feg}_{ }
 \bigl( \bar d^{f}_{ } C\gamma_5^{ }
 \bar u^{g \rm{T}}_{ } \bigr)
 \bar d_\alpha^e 
 \,,
\end{align}
where $a,b,c$ are colour indices and $C$ is the charge-conjugation
operator, defined in appendix~\ref{app:nrqcd}. The contraction with the
totally anti-symmetric symbol $e^{abc}$ guarantees  
gauge invariance. The
two-point correlation functions considered are
\begin{align}
\label{eq:Ncorr}
    {\cal C}_{\pm}^{ }(x_3^{ }) \; \equiv \;  
    \int_0^{1/T} \! {\rm d} x_0^{ }
    \, e^{-i k_0 x_0}_{ } \int_{\mathbf{r}}
   \Tr \expval{N(x_0^{ },x^{ }) \, \overline{\!N}(0) P_{\pm}^{ }} 
   \, , \quad
   x \;\equiv\; (\mathbf{r},x^{ }_3) 
   \, ,
\end{align}
where $k_0^{ } = \pi T$ is the lowest positive fermionic Matsubara frequency,
arising due to anti-periodic boundary conditions in the temporal
direction, $P_\pm^{ }=\left(1\pm\gamma_3^{ }\right)/2$ 
is the $x_3^{ }$-parity
projector, the trace is over Dirac space, 
and $\int^{ }_{\mathbf{r}} \equiv \int \! {\rm d}^2\mathbf{r}$ 
denotes an integration over the transverse spatial directions. 
The corresponding screening
masses, characterizing the long-distance behaviour of
eq.~\eqref{eq:Ncorr}, are defined as
\begin{align}
    m_\pm^{ } \; \equiv \; -\lim_{x_3\to \infty} \frac{{\rm d}}{{\rm d} x_3}
 \ln \left[ {\cal C}_\pm^{ } (x_3^{ }) \right] \, .
\end{align}

In vacuum, due to the spontaneous breaking of
chiral symmetry, the positive 
($N$, ${\cal C}_{+}^{ }$) and the negative ($N^*_{ }$, ${\cal C}_{-}^{ }$) 
parity partners 
have masses which differ by several hundreds
of MeV~\cite{Workman:2022ynf}. On the contrary, at high temperatures,
thanks to the effective restoration of chiral symmetry, 
the screening masses associated with the parity
partners  are expected to become degenerate 
(for numerical evidence, see
refs.~\cite{Datta:2012fz,Rohrhofer:2019yko,Rescigno:2023tkq}).

%
\section{Low-energy (long-distance) description}
\label{sec:eft}

%
\subsection{Effective action}

In QCD at very high temperatures, adopting the imaginary-time formalism, 
field fluctuations which depend on the temporal coordinate carry 
a large ``mass'', and should therefore represent weakly coupled 
short-distance physics. 
The strongly coupled long-distance degrees of freedom are given 
by the zero Matsubara modes of the gauge fields. Given that quarks
obey antiperiodic boundary conditions over the temporal direction, 
they are always ``heavy'', even if their vacuum mass would vanish. 

The gluonic sector of the effective theory contains
gauge fields $A_k^{ }$, with $k=1,2,3$, living in three spatial
dimensions, whose dynamics
is non-perturbative \cite{Linde:1980ts}. They are coupled to a
massive scalar field $A_0^{ }$, which transforms under the adjoint
representation of the gauge group
(cf.\ appendix~\ref{app:eft} for a more detailed discussion). 
By taking into account these degrees of freedom,
the corresponding effective action, which is usually called
Electrostatic QCD (EQCD), reads
\cite{Ginsparg:1980ef,Appelquist:1981vg}
\begin{align}
\label{eq:EQCD}
    S_\rmii{EQCD}^{ } \, = \, \int\! {\rm d}^3_{ }x\,
   \lb \frac{1}{2}
   \Tr \left[ F_{ij}^{ }F_{ij}^{ } \right] 
  +\Tr\left[\left(D_j^{ } A_0^{ } \right)\left(D_j^{ } A_0^{ } \right) \right]
  + m_\rmii{E}^2 \Tr \left[ A_0^2 \right]\rb \, + \, \dots \, ,
\end{align}
where the dots stand for higher-dimensional operators
\cite{Laine:2016hma}. Here $i,j=1,2,3$ and 
$
 [ D_i^{ },D_j^{ } ]=-i g_\rmii{E}^{ } F_{ij}^{ }
$ with the covariant derivative defined as
$
 D_i^{ }=\partial_i^{ }-ig_\rmii{E}^{ }A_i^{ }
$.  
The matching coefficients 
$m^2_\rmii{E}$ and $g^2_\rmii{E}$ parametrize the Lagrangian
mass squared of the scalar field $A_0^{ }$ and
the dimensionful coupling constant of the three-dimensional Yang-Mills
theory, respectively. They have been matched to QCD at several orders in
perturbation theory, with the leading-order 
expressions reading
\cite{Kapusta:1979fh,Laine:2005ai,Ghisoiu:2015uza}
\be
    m^2_\rmii{E} \, = \, g^2_{ }T^2_{ }
    \left(1+\frac{N^{ }_\rmii{\sl f}}{6}\right)
    + O(g^4_{ }T^2_{ })
    \,, \qquad
    g^2_\rmii{E} \, =\, g^2_{ } T 
    + O(g^4_{ } T)
   \, , \label{eq:ge}
\ee
where $g$ is the QCD coupling constant 
and $N_\rmi{\sl f}^{ }\,$ is the number of flavours.  

On the other hand, at high temperatures quarks are
heavy with masses $\sim \pi T$, due to fermionic Matsubara
frequencies. The dynamics of such fields is described by
three-dimensional non-relativistic QCD 
(NRQCD)~\cite{Huang:1995tz,Caswell:1985ui,Brambilla:1999xf}, see
appendix~\ref{app:nrqcd} for a derivation. 
By taking into account the power-counting rules in 
table~\ref{tab:power_counting} of 
appendix~\ref{app:nrqcd}, which imply that normal rather than
covariant derivatives can be used in the transverse directions, 
the effective action for $N_\rmi{\sl f}^{ }=3$ massless
fermions in the lowest Matsubara sectors reads
\begin{align}
\label{eq:NRQCD}
    \begin{aligned}
        S_\rmii{NRQCD}^{ }  = i \sum_\rmi{\sl f\,=\it u,d,s}
 \int\! {\rm d}^3_{ } x \, \biggl\{
 & \bar{\chi}_\rmi{\sl f\,}^{ }(x) 
 \left[ M-g_\rmii{E}^{ } A_0^{ } +D_3^{ } 
 -\frac{\nabla_\perp^2}{2\pi T}\right]\chi_\rmi{\sl f\,}^{ }(x) \\
 -& \bar{\phi}_\rmi{\sl f\,}^{ }(x) \left[ M+g_\rmii{E}^{ } A_0^{ } +D_3^{ }
 -\frac{\nabla_\perp^2}{2\pi T} \right]\phi_\rmi{\sl f\,}^{ }(x) \biggr\}
 \, + O\left( \frac{g_\rmii{E}^2}{\pi T} \right) ,
    \end{aligned}
\end{align}
where $\chi$ and $\phi$ are two-component
spinors defined in the paragraph below eq.~\eqref{eq:dirac3},\footnote{%
 While $\bar\chi^{ }_\rmi{\sl f}$ and $\chi^{ }_\rmi{\sl f}$ are
 independent variables in the path-integral formulation, in the 
 canonical formulation they are related through  
 $\bar\chi^{ }_\rmi{\sl f} = \chi_\rmi{\sl f}^\dagger\,$, 
 and similarly for $\phi_\rmi{\sl f}^{ }\,$. \label{fn:bar} 
 }
 {\sl f\,} is a flavour index, and
\begin{align}
\label{eq:M}
    M \, = \, \pi T \left(1+\frac{g^2}{6\pi^2}\right)
    + O(g^4_{ } T)
\end{align}
is a matching coefficient that was computed 
at 1-loop order in ref.~\cite{Laine:2003bd}.  
In eq.~\eqref{eq:NRQCD}, 
according to the power-counting rules, 
we neglected higher-dimensional operators, see
ref.~\cite{Huang:1995tz} for some of them. Note that the action in
eq.~(\ref{eq:NRQCD}) displays more symmetries than the full theory, 
with e.g.\ $\chi_\rmi{\sl f\,}^{ }$ and $\phi_\rmi{\sl f\,}^{ }$ being
independent of each other at this order.
The QCD dynamics at high temperature is now described by 
$S_{\rmii{QCD}_3}^{ } = S_\rmii{EQCD}^{ } + S_\rmii{NRQCD}^{ }$.

%
\subsection{Equations of motion}
\label{sec:EOM}

From the effective action in eq.~\eqref{eq:NRQCD} and an 
infinitesimal transformation of path-integration variables, it is
straightforward to see that
the three-dimensional fields $\chi$ and $\phi$ for
each flavour\hspace*{0.1mm}\footnote{%
 The flavour index is omitted unless
 it is necessary for the clarity
 of presentation.}  
satisfy for a generic
interpolating operator $O(y)$ the equations of motion
\begin{align}
\label{eq:EOM_chi}
    i \expval{\left[ M - g_\rmii{E}^{ } A_0^{ } + D_3^{ }
 - \frac{\nabla_\perp^2}{2\pi T} \right] \chi(x)\, O(y)} \;
 &=\;\;\; \expval{\frac{\delta O(y)}{\delta \bar{\chi} (x)}}
 \, ,  \\[0.25cm]
    i \expval{\left[ M + g_\rmii{E}^{ } A_0^{ } + D_3^{ }
 - \frac{\nabla_\perp^2}{2\pi T}\right] \phi(x)\, O(y)} \;
 &= - \expval{\frac{\delta O(y)}{\delta \bar{\phi} (x)}} \, ,
\label{eq:EOM_phi}
\end{align}
where the derivatives act on the $x$ coordinates. Analogous
equations hold for $\bar{\chi}$ and $\bar{\phi}$.  
The propagators of $\chi$ and $\phi$ are defined as
\begin{align}
    \label{eq:prop}
    S_{\chi} (x) \;\equiv\;
   \expval{\chi(x)\,\bar{\chi}(0)}_{\rm f}\, ,\qquad     
    S_\phi (x)   \;\equiv\;
   \expval{\phi(x)\,\bar{\phi}(0)}_{\rm f} \, ,     
\end{align}
where in eq.~(\ref{eq:prop}) the expectation value $\expval{\cdot}_{\rm f}$
indicates the path integral over fermions only.  
Then by choosing $O =\bar{\chi}$, $\bar{\phi}$ at $y=0$ 
in eqs.~\eqref{eq:EOM_chi} and \eqref{eq:EOM_phi}, respectively, 
the fermion propagators satisfy the equations
\begin{align}
\label{eq:S_chi}
 & \expval{\left[ M + \partial_3^{ }   - \frac{\nabla_\perp^2}{2\pi T}\right]
 S_{\chi}^{ } (x)}= 
 g_\rmii{E}^{ } \expval{\Big[i A_3^{ }(x) + A_0^{ }(x)\Big]
 S_{\chi}^{ } (x)} - i \id \delta^{(3)}_{ }(x)\, ,\\[0.125cm]
 & \expval{\left[ M + \partial_3^{ }
  - \frac{\nabla_\perp^2}{2\pi T}\right] S_\phi^{ } (x)}=
 g_\rmii{E}^{ } \expval{\Big[i A_3^{ }(x) - A_0^{ }(x)\Big] S_\phi^{ } (x)}
 + i \id \delta^{(3)}_{ }(x) \, ,
\label{eq:S_phi}
\end{align}
where $\id$
stands for the identity in spinor and colour indices.  
Since the fermions had been integrated out, 
the expectation values in eqs.~(\ref{eq:S_chi}) and (\ref{eq:S_phi})
indicate the path integral over the gauge fields. 
Note that these equations are valid
also without integrating over the gauge fields,
i.e.\  for a fixed gauge field background, and that
at this order the fermion
propagators are diagonal in flavour and spin.

%
\subsection{Perturbation theory}

The free fermion action is obtained by setting $g_\rmii{E}^{ }=0$ in
eq.~\eqref{eq:NRQCD}. 
The corresponding equations of motion 
are readily worked out from eqs.~\eqref{eq:S_chi}
and \eqref{eq:S_phi}. 
The free propagators can be written as
(the coordinate-space expression is given
in eq.~\eqref{eq:prop_free_coord})
\begin{align}
    \label{eq:free_prop1}
    S_{\chi}^{(0)} (\mathbf{r}, x_3^{ }) 
  &= -i  \theta(x_3^{ }) \id\!\! \int_{\textbf{p}}
  e^{i \textbf{p} \cdot \textbf{r}  }\, 
   e^{-x_3 \bigl( M +\frac{\textbf{p}^2}{2 \pi T} \bigr)} \, ,\\[0.25cm]
 S^{(0)}_\phi (\mathbf{r}, x_3^{ }) 
 &= - \, S^{(0)}_\chi (\mathbf{r}, x_3^{ })\, ,     
    \label{eq:free_prop2}
\end{align}
where $\int_{\mathbf{p}}\equiv \int\! {\rm d }^2_{ } \mathbf{p}/(2\pi)^2_{ }$.
At next-to-leading order in $g_\rmii{E}^{ }$, we can define
\be
 S_{\chi}^{ } (\mathbf{r}, x_3^{ }) =
  S_{\chi}^{(0)} (\mathbf{r}, x_3^{ })
 + g_\rmii{E}^{ }\, S_{\chi}^{(1)} (\mathbf{r}, x_3^{ }) 
 + O(g_\rmii{E}^2 )\; , 
\ee
and analogously for $S_{\phi}^{ } (\mathbf{r}, x_3^{ })$. 
By solving eqs.~(\ref{eq:S_chi}) and (\ref{eq:S_phi}) and 
approximating the transverse movement 
(cf.\ appendix~\ref{app:prop}), we obtain
\begin{align}
\label{eq:ge_prop1}
 S_{\chi}^{(1)} (\mathbf{r}, x_3^{ })
 & \,\simeq\, \int_0^{x_3}\! {\rm d}z^{ }_3\, \big[i A_3^{ } + A_0^{ }\big]
 \Big(\frac{z_3}{x_3}\textbf{r},z_3^{ }\Big)
 \, S_{\chi}^{(0)} (\mathbf{r}, x_3^{ })
 \,, \\
 S_{\phi}^{(1)} (\mathbf{r}, x_3^{ })
 & \,\simeq\, \int_0^{x_3}\! {\rm d}z^{ }_3\, \big[i A_3^{ } - A_0^{ } \big]
\Big(\frac{z_3}{x_3}\textbf{r},z_3^{ }\Big)\,
 S_{\phi}^{(0)} (\mathbf{r}, x_3^{ })
 \,.
\label{eq:ge_prop2}
\end{align} 

%
\section{Baryonic correlators in the effective theory}
\label{sec:bar}

%
\subsection{Contractions}

The effective-theory expression 
for the baryonic interpolating operator from
eq.~\eqref{eq:nucleon} is readily
obtained by using the definitions in appendix~\ref{app:nrqcd}. After taking
the Fourier transform in the time direction, we may restrict to the 
contributions that involve the lowest Matsubara modes propagating
in the positive $x^{ }_3$-direction. This implies that $N$ is 
represented by two $\chi$ fields and one $\phi$ field. 
Furthermore, by displacing the fundamental fields in the transverse
direction, i.e.\  by introducing a point-splitting,
the Fourier transform in the compact direction
of the baryonic interpolating operator leads to
\bea
 N^{ } (\mathbf{r}_1^{ },\mathbf{r}_2^{ },\mathbf{r}_3^{ };x_3^{ })
 & \to &
 \epsilon^{abc}_{ }\, 
 \big[\, 
 \chi^{aT}_{u} (\mathbf{r}_1^{ },x_3^{ })
 \,\sigma_2^{ }\, \phi_{d}^b(\mathbf{r}_2^{ },x_3^{ })
 + \phi^{aT}_{u}(\mathbf{r}_1^{ },x_3^{ })
 \,\sigma_2^{ }\, \chi_{d}^b(\mathbf{r}_2^{ },x_3^{ })
 \,\big]
 \, \chi^{c}_{d,\alpha}(\mathbf{r}_3^{ },x_3^{ })
 \,, \nonumber \\[0.25cm]
 \overline{\!N}^{ } (0)
 & \to &  
 \epsilon^{feg}_{ }\,
 \big[\, 
 \bar\phi^{f}_{d}({0}^{ })
 \,\sigma_2^{ }\,
 \bar\chi_{u}^{gT}({0}^{ })
  + 
 \bar\chi^{f}_{d}({0}^{ })
 \,\sigma_2^{ }\, 
 \bar\phi^{g T}_{u}({0}^{ })
 \,\big]
 \, 
 \bar\chi^{e}_{d,\alpha}({0}^{ }) 
  \,,
 \label{eq:N_FT}
\eea
where $\alpha$ is a two-component spinor index. 
To avoid clutter, we have omitted overall factors $T^{3/2}_{ }$ from 
both operators, originating from eq.~\eqref{eq:chiphi},  
however they are restored in eq.~\eqref{eq:Ncorr_eff_wick}. 
The two-point correlators from eq.~\eqref{eq:Ncorr} are defined
in the effective theory as
\bea
    {\cal C}_{\pm}^{ }(x_3^{ }) & = & 
    \int_{\mathbf{r}} 
    {\cal C}_{\pm}^{ }
    (\textbf{r}^{ },\textbf{r}^{ },\textbf{r}^{ };x_3^{ })
   \;, 
   \label{eq:Cpm_eff}
   \\[2mm] 
   {\cal C}_{\pm}^{ }
   (\textbf{r}_1^{ },\textbf{r}_2^{ },\textbf{r}_3^{ };x_3^{ })
   & \equiv & 
   \frac{1}{T} \, 
   \Tr \expval{N(\textbf{r}_1^{ },\textbf{r}_2^{ },\textbf{r}_3^{ };x_3^{ })
    \,\overline{\!N}(0) P_{\pm}^{ }} \,,
  \label{eq:Ncorr_eff}
\eea
where $1/T$ comes from $\int_0^{1/T}\!{\rm d}x^{ }_0$,  
and
$
 P_{\pm}^{ } 
 =
 (\pm i/2) \id^{ }_{ }
 \stackrel{\rmii{\eqref{eq:dirac}}}{\leftrightarrow} 
 [\gamma^{ }_0 
 (\id^{ }_{ } \pm \gamma^{ }_3)/2]^{ }_{11}
$. 
We have used the
same symbols as in QCD, 
given that the ambiguity can be resolved from the context. 

We remark that, as long as 
$\mathbf{r}^{ }_i \neq \mathbf{r}^{ }_j$, 
neither the interpolating operator above eq.~\eqref{eq:N_FT}
nor the correlation function in eq.~\eqref{eq:Ncorr_eff} is gauge
invariant under gauge transformations involving the transverse
coordinates. Gauge invariance could be restored 
by contracting the point-split operator 
with transverse Wilson lines. However, such
transverse Wilson lines play no r\^ole in the calculation of the
screening masses. Indeed, the final result will be
gauge independent even without them, 
as gauge dependence vanishes in 
the large-separation limit in the longitudinal direction.\footnote{%
 The technical reason is that we need the component 
 $\Delta^{ }_{33}$ of the gauge propagator, 
 cf.\ eq.~\eqref{eq:calVpm}, but the third component 
 of the momentum is zero, cf.\ eq.~\eqref{eq:large_x_3}, 
 so only the transverse part plays a r\^ole.
 }
In order to streamline 
the presentation, we omit the transverse Wilson lines, 
and display results
only for the Feynman gauge.

By exploiting the antisymmetry of the Levi-Civita symbol, 
and noting that the propagators are flavour independent, 
integration over the fermionic fields yields
\begin{align}
\label{eq:Ncorr_eff_wick}
    {\cal C}_{\pm}^{ }
  (\textbf{r}_1^{ },\textbf{r}_2^{ },\textbf{r}_3^{ };x_3^{ })
  = \mp \, T^2_{ } \,\Big\langle\,
   2 W(\textbf{r}_1^{ },\textbf{r}_2^{ },\textbf{r}_3^{ };x_3^{ })
 + 3 W(\textbf{r}_2^{ },\textbf{r}_1^{ },\textbf{r}_3^{ };x_3^{ }) 
        \,\Big\rangle \,,
\end{align}
where the Wick contraction is defined as
\begin{align}
 \label{eq:Wick}
 W(\textbf{r}_1^{ },\textbf{r}_2^{ },\textbf{r}_3^{ };x_3^{ })
  \; \equiv \; 
 -i  \, \epsilon^{abc}_{ } \epsilon^{gfe}_{ }\, 
 S^{ag}_{\chi} (\mathbf{r}_1^{ }, x_3^{ })\,
 S^{bf}_{\phi} (\mathbf{r}_2^{ }, x_3^{ }) 
 S^{ce}_{\chi} (\mathbf{r}_3^{ }, x_3^{ })\; .
\end{align}

Equation~\eqref{eq:Ncorr_eff_wick} implies that our ${\cal C}_{\pm}^{ }$
is a sum of two independent correlation functions. 
This is a consequence of the fact that the action 
in \eqref{eq:NRQCD} displays ``emergent'' global symmetries, 
with the numbers of $\chi$ and $\phi$-particles 
separately conserved. Given that the two correlators 
in \eqref{eq:Ncorr_eff_wick} differ just by a permutation
of coordinates, and that in the end all coordinates are set equal
(cf.\ eq.~\eqref{eq:Cpm_eff}), the two correlators 
yield the same baryonic screening mass. 
This degeneracy could be 
broken by higher-dimensional 
operators in the effective theory~\cite{Huang:1995tz}, leading to 
a ``fine structure'' of the screening spectrum, but this is an 
effect of higher order than our $O(g^2_{ } T)$.   

%
\subsection{Free limit}

By inserting the free propagators from eqs.~\eqref{eq:free_prop1} 
and \eqref{eq:free_prop2} into eq.~\eqref{eq:Wick}, we see that 
\begin{align}
\label{eq:leading_prop}
 W^{(0)}_{ }(\textbf{r}_1^{ },\textbf{r}_2^{ },\textbf{r}_3^{ };x_3^{ })  = 
  - 6\, \theta(x_3^{ })
 \int_{\textbf{p}_1,\textbf{p}_2,\textbf{p}_3}
 e^{i (\textbf{p}_1^{ } \cdot \textbf{r}_1^{ }
  + \textbf{p}_2^{ } \cdot \textbf{r}_2^{ } 
  +\textbf{p}_3^{ } \cdot \textbf{r}_3)^{ } }
  e^{-x_3^{ }\big(3M + \frac{\textbf{p}^2_1}{2\pi T}
 +\frac{\textbf{p}^2_2}{2\pi T}+\frac{\textbf{p}^2_3}{2\pi T}\big)}
 \,.
\end{align}
{}From here it follows that $W^{(0)}_{ }$
satisfies a (2+1)-dimensional Schr\"odinger equation,  
\begin{align}
\label{eq:EOM:tree}
 \left[ 3M + \partial_3^{ } -\sum_{i=1}^3 
 \frac{\nabla^2_{\mathbf{r_i}}}{2\pi T}\right]
 W^{(0)}_{ } (\textbf{r}_1^{ },\textbf{r}_2^{ },\textbf{r}_3^{ };x_3^{ })
 \, \stackrel{x^{ }_3 > 0 }{=} \, 0 \, .
\end{align}
Thus, if quarks have small transverse momentum
(indeed we will see that parametrically 
$
 \nabla^2_{\mathbf{r_i}} \sim m_\rmii{E}^2 \sim g^2_{ }T^2_{ }
$), 
the exponential falloff is
dominated by $3M =  3\pi T + O(g^2_{ })$, 
which then represents the leading contribution
to the baryonic screening masses $m^{ }_\pm$. 


%
\subsection{Next-to-leading order}
\label{sec:eom_bar}

To date, the only estimate of $O(g^2_{ })$ corrections to 
a baryonic screening mass in the high temperature regime of QCD is
qualitative \cite{Hansson:1994nb}. Here the full $O(g^2_{ })$
correction is derived in 
the same way as for the mesonic case \cite{Laine:2003bd}, 
demonstrating in particular its IR finiteness
up to this order in the weak-coupling expansion.

The equation of motion for $W$ in the interacting case is readily
worked out from eqs.~(\ref{eq:S_chi}) and (\ref{eq:S_phi}), and it
reads
\begin{align}
\label{eq:EOMge}
 & \hspace{-0.5cm} \biggl[ 3M + \partial_3^{ } \! -\! \sum_{i=1}^3 
  \frac{\nabla^2_{\mathbf{r_i}}}{2 \pi T} \biggr]
 \!\big\langle W(r_1^{ },r_2^{ },r_3^{ })\big\rangle
 \stackrel{x_3^{ }> 0}{=}
 -i g_\rmii{E}^{ } 
 \epsilon^{abc}_{ } \epsilon^{gfe}_{ }
  \bigg\langle \Big[(i A_3^{ }\! +\! A_0^{ }) S_{\chi}^{ }\Big]^{ag}_{ }
 \!\! (r_1^{ }) S^{bf}_{\phi} (r_2^{ })
 S^{ce}_{\chi} (r_3^{ }) \nonumber\\[0.125cm]
 & + S_{\chi}^{ag}(r_1^{ }) \Big[(i A_3^{ }\! -\! A_0^{ })
  S_{\phi}^{ }\Big]^{bf}\!\! (r_2^{ }) S^{ce}_{\chi} (r_3^{ }) +
  S_{\chi}^{ag}(r_1) S^{bf}_{\phi}(r_2)\Big[(i A_3\! +\! A_0)
  S_{\chi}\Big]^{ce}_{ }\!\! (r_3^{ })\bigg\rangle\, ,
\end{align}
where we have introduced $r_i^{ } \equiv (\textbf{r}_i^{ },x_3^{ })$ 
to simplify the notation. Inserting $S^{ }_{\chi}$ and 
$S^{ }_{\phi}$ from 
eqs.~\eqref{eq:ge_prop1} and \eqref{eq:ge_prop2}, respectively,  
and performing the gluon contractions 
(cf.\ appendix~\ref{app:int} for the gluon propagator and further
details on intermediate steps), we get
\begin{align}\label{eq:EOMW}
  \biggl[ 3M + \partial_3^{ } \! -\!\sum_{i=1}^3 
 \frac{\nabla^2_{\mathbf{r_i}}}{2 \pi T} \biggr]
  \big\langle W(r_1^{ },r_2^{ },r_3^{ })\big\rangle 
 \stackrel{x_3^{ }> 0}{=} 
 -\, {\cal U}(r_1^{ },r_2^{ },r_3^{ })\, 
 W^{(0)}_{ }(r_1^{ },r_2^{ },r_3^{ })
 + O(g_\rmii{E}^3)
 \,. 
\end{align}
Here, with the notation from eq.~\eqref{eq:calVpm}, 
\bea
 && \vspace*{-1.5cm}
   {\cal U}(r_1^{ },r_2^{ },r_3^{ })  =  
   \frac{4 g_\rmii{E}^2 }{3} 
 \bigg\{\,
   \frac{ 
     {\cal V}^{-}_{ }(\textbf{r}_1^{ },\textbf{r}_2^{ },x_3^{ })
   + {\cal V}^{-}_{ }(\textbf{r}_2^{ },\textbf{r}_1^{ },x_3^{ })
   }{2}  
   + \frac{
     {\cal V}^{+}_{ }(\textbf{r}_1^{ },\textbf{r}_3^{ },x_3^{ })
   + {\cal V}^{+}_{ }(\textbf{r}_3^{ },\textbf{r}_1^{ },x_3^{ })
   }{2}  
  \nonumber\\
  && 
   + 
   \, \frac{
   {\cal V}^{-}_{ }(\textbf{r}_2^{ },\textbf{r}_3^{ },x_3^{ })
   + {\cal V}^{-}_{ }(\textbf{r}_3^{ },\textbf{r}_2^{ },x_3^{ }) }{2}
   - {\cal V}^{+}_{ }(\textbf{r}_1^{ },\textbf{r}_1^{ },x_3^{ })
   - {\cal V}^{+}_{ }(\textbf{r}_2^{ },\textbf{r}_2^{ },x_3^{ })
   - {\cal V}^{+}_{ }(\textbf{r}_3^{ },\textbf{r}_3^{ },x_3^{ })
 \,\bigg\} \,. \nonumber \\ 
 \label{eq:Kstor}
\eea
To extract the
screening masses, 
we take the limit $x_3^{ }\rightarrow\infty$,
which leads to
\begin{align}
 U(\textbf{r}_1^{ },\textbf{r}_2^{ },\textbf{r}_3^{ })
 \; \equiv \;
 \lim_{x_3\rightarrow\infty} {\cal U}(r_1^{ },r_2^{ },r_3^{ }) =
 \frac{1}{2}\Big[\,
  V^{-}(r_{12}^{ }) + V^{+}(r_{13}^{ }) + V^{-}(r_{23}^{ })
 \,\Big] 
 \,, \label{eq:U}
\end{align}
where 
$r_{ij}^{ } \equiv |\textbf{r}_i^{ } - \textbf{r}_j^{ }|$, 
and  $V^\pm_{ }$ are the static potentials defined in
ref.~\cite{Brandt:2014uda}, 
\begin{align}
  V^\pm_{ } (r) \; & \equiv\; \frac{4}{3} \frac{g_\rmii{E}^2}{2\pi}
  \biggl[ \ln \Bigl( {\frac{m_\rmii{E} r}{2} \Bigr) + \gamma_\rmii{E}
 \pm K_0^{ }(m_\rmii{E}^{ } r)} \biggr]
 \,,  \label{eq:Vpm}
\end{align}
where 
$\gamma_\rmii{E}^{ }$ 
is the Euler-Mascheroni constant
and 
$K_0^{ }$ is a modified Bessel function. 
It is appropriate to stress that according to eq.~\eqref{eq:U}, 
the three-body potential receives contributions from 
two-body interactions only. 

Finally, by replacing
$W^{(0)}_{ }\rightarrow \langle W \rangle$ on the 
right-hand side of eq.~(\ref{eq:EOMW}), 
which is justified at $O(g_\rmii{E}^2)$ and implements a resummation
of potential-like interactions, 
and taking the already-mentioned limit $x_3^{ }\rightarrow\infty$, 
the equation of motion for the correlator reads
\begin{align}
 \label{eq:EOM-full}
 \left[\partial_3^{ } - \sum_{i=1}^3 
 \frac{\nabla^2_{\mathbf{r_i}}}{2 \pi T}
 + V(\textbf{r}_1^{ },\textbf{r}_2^{ },\textbf{r}_3^{ }) \right]
   \big\langle\, W(\textbf{r}_1^{ },\textbf{r}_2^{ },\textbf{r}_3^{ };
   x_3^{ })
   \,\big\rangle  \; \stackrel{ }{=}  \; 0 + O(g_\rmii{E}^3) 
   \, ,
\end{align}
where 
\begin{align}
\label{eq:pot}
  V(\textbf{r}_1^{ },\textbf{r}_2^{ },\textbf{r}_3^{ }) 
  \; \equiv \; 
  3 M + 
  U(\textbf{r}_1^{ },\textbf{r}_2^{ },\textbf{r}_3^{ })
  \, .
\end{align}
The two contributions in eq.~\eqref{eq:Ncorr_eff_wick} satisfy 
the same equation, just with a permutation of coordinates
(which in the end are set the same, cf.\ eq.~\eqref{eq:Cpm_eff}). 
Therefore, the solutions of both equations
yield the same screening mass, which is then also 
the screening mass extracted from 
${\cal C}^{ }_{\pm}(x_3^{ })$ at $O(g_\rmii{E}^2)$.

We end this section by remarking that the potential 
$U(\textbf{r}_1^{ },\textbf{r}_2^{ },\textbf{r}_3^{ })$ from 
eq.~\eqref{eq:U} is symmetric in the exchange of its first and
last coordinate. For the first contribution from 
eq.~\eqref{eq:Ncorr_eff_wick}, this corresponds to 
$
 \textbf{r}_1^{ } \leftrightarrow \textbf{r}_3^{ }
$, 
which in terms of eq.~\eqref{eq:N_FT} is due to  
an accidental symmetry in the action from the exchange 
$\chi^{ }_u \leftrightarrow \chi^{ }_d$.
In contrast, 
for the second contribution from 
eq.~\eqref{eq:Ncorr_eff_wick}, this corresponds to 
$
 \textbf{r}_2^{ } \leftrightarrow \textbf{r}_3^{ }
$, 
the exchange of two identical $\chi^{ }_d$ particles.

%
\section{Schr\"odinger equation for baryonic correlators}
\label{sec:schrodinger}

%
\subsection{Center-of-mass coordinates}

From the discussion in sec.~\ref{sec:eom_bar}, for large separations
in the longitudinal direction, the equation of motion for a generic
two-point correlation function related to baryonic interpolating
operators (in both parity channels) implies the Schr\"odinger equation
\begin{align}
    \bigg[\, -\frac{\nabla_{\mathbf{r}_1}^2
     +\nabla_{\mathbf{r}_2}^2
     +\nabla_{\mathbf{r}_3}^2}{2\pi T}
  + V(\textbf{r}_1^{ },\textbf{r}_2^{ },\textbf{r}_3^{ })
    \,\bigg]\,
  \psi\left( \mathbf{r}_1^{ },\mathbf{r}_2^{ },\mathbf{r}_3^{ } \right)
 \, = E\, \psi\left( \mathbf{r}_1^{ },
 \mathbf{r}_2^{ },\mathbf{r}_3^{ } \right) \, ,
\end{align}
where the potential is given by eq.~(\ref{eq:pot}). The energy
eigenvalue of the ground state yields our screening masses, 
i.e.\ $m^{ }_\pm = \min\{E\} + O(g^3_{ } T)$.

In order to solve the three-body Schr\"odinger equation, 
we employ the Jacobi coordinates
\begin{align}
    \begin{aligned}
 \label{eq:Jacobi}
        \mathbf{R} \; & \equiv \;
   \frac{\mathbf{r}_1+\mathbf{r}_2+\mathbf{r}_3}{3}
  \,, \\
        \bsxi_1^{ } \; & \equiv \;
    \mathbf{r}_3^{ } - \mathbf{r}_1^{ }
  \,, \\
        \bsxi_2^{ } \; & \equiv \;
  \frac{2}{\sqrt{3}}
  \left( \mathbf{r}_2^{ }
 -\frac{ \mathbf{r}_1+\mathbf{r}_3  }{2} 
  \right)
  \, = \, \sqrt{3}\left( \mathbf{r}_2^{ }-\mathbf{R}\right)
 \, ,
 \end{aligned}
\end{align}
where $\textbf{R}$ is the position of the center-of-mass in the
transverse plane, $\bsxi_1^{ }$ is the relative separation between two
quarks located at $\mathbf{r}_1^{ }$ and $\mathbf{r}_3^{ }$,
while $\bsxi_2^{ }$ describes, up to some numerical factor, the relative
separation between the quark in $\mathbf{r}_2^{ }$ and the
center-of-mass of the other pair. In this sense the set of coordinates
$\left(\bsxi_1^{ }, \bsxi_2^{ } \right)$ describes the relative separations of
the underlying two-body problems. With this change of
variables the potential only depends on
$\bsxi_1^{ }$ and $\bsxi_2^{ }$, given that
\begin{align}
\begin{aligned}
    \mathbf{r}_2^{ }-\mathbf{r}_1^{ }
 & = \frac{\bsxi_1^{ } + \sqrt{3} \bsxi_2^{ }}{2} 
 \,, \\
    \mathbf{r}_3^{ }-\mathbf{r}_2^{ }
 & = \frac{\bsxi_1^{ }-\sqrt{3} \bsxi_2^{ }}{2} 
 \, .
 \label{eq:xis}
\end{aligned}
\end{align}
In this way, the Laplace operator can be separated into the
center-of-mass and relative motions. The
Schr\"odinger equation for the relative motion can be written as
(cf.\ eq.~\eqref{eq:laplacian})
\begin{align}
\label{eq:schr}
    \left[ - \frac{1}{\pi T}
   \left( \nabla_{\bsxi_1}^2+\nabla_{\bsxi_2}^2  \right)
   \, + \, V \left( \bsxi_1^{ }, \bsxi_2^{ } \right)\right]
  \psi\left(\bsxi_1^{ },\bsxi_2^{ } \right) 
 \, = \, E\, \psi(\bsxi_1^{ },\bsxi_2^{ }) \, ,
\end{align}
where it is understood that the static potential and the wave function
are expressed in terms of the new coordinates 
$\bsxi_1^{ }$ and $\bsxi_2^{ }$.

%
\subsection{Numerical solution}

In order to find a numerical solution to eq.~\eqref{eq:schr}, 
it is convenient to define the dimensionless transverse coordinates
\begin{align}
    \hat{\bsxi}_1^{ } \, \equiv \, m_\rmii{E}^{ } \, \bsxi_1^{ }
  \,, \quad \hat{\bsxi}_2^{ } \, \equiv m_\rmii{E}^{ } \, \bsxi_2^{ } 
  \,.
\label{eq:dimensionless}
\end{align}
Moreover, we express $E$ in terms of a dimensionless 
eigenvalue $\hat{E}$, by writing
\begin{align}
    E \; = \; 3 M + \frac{4}{3}\frac{g^2_{ } T}{2 \pi }\hat{E}
   + O(g^3_{ }T)
   \; \stackrel{\rmii{\eqref{eq:M}}}{=} \; 
   3\pi T + \frac{g^2 T}{2\pi} \biggl( 1 +\frac{4\hat{E}}{3} \biggr)
    + O(g^3_{ }T) \,.
   \label{eq:Ehat_def}
\end{align}
This leads to a Schr\"odinger equation
in terms of dimensionless variables, 
\begin{align}
\label{eq:schrodinger}
    \left[ -\frac{1}{\rho}\left({\nabla^2_{\hat{\bsxi}_1}}
 +{\nabla^2_{\hat{\bsxi}_2}}\right) +\hat{V}(\hat{\bsxi}_1,\hat{\bsxi}_2)
 - \hat{E} \right] \, \psi (\hat{\bsxi}_1^{ },\hat{\bsxi}_2^{ })\, = \, 0 \, ,
\end{align}
where $\hat{V}$ is a rescaled static potential
from eqs.~\eqref{eq:U} and \eqref{eq:Vpm},
\begin{align}
  \hat{V} \; \equiv \; 
  \frac{
     \hat{V}^-_{ }(\hat{r}^{ }_{12})
   + \hat{V}^+_{ }(\hat{r}^{ }_{13})
   + \hat{V}^-_{ }(\hat{r}^{ }_{23})
       }{2}
   \;, \quad 
    \hat{V}^\pm (\hat{r}) 
   \; \equiv \; \ln \frac{\hat{r}}{2}
  + \gamma_\rmii{E}^{ } \pm K_0^{ }(\hat{r})
  \,,
\end{align}
and $\rho$ is a re-parametrization of the dimensionful quantities
of the problem~\cite{Laine:2003bd},
\begin{align}
    \rho \, \equiv \, \frac{4}{3}\frac{g^2_{ } T}{2\pi}
  \frac{\pi T}{m_\rmii{E}^2} 
  \, \overset{N_\rmii{\sl f}\,=\,3}{=} \, \frac{4}{9} 
  + O(g^2_{ })
  \, .
\end{align}

Equation~\eqref{eq:schrodinger} can be solved numerically by exploiting 
a two-dimensional generalization of the so-called hyperspherical
harmonics method. It is usually employed for 
three-dimensional quantum many-body problems, see
ref.~\cite{das2015hyperspherical} for an introduction and
appendix~\ref{app:HH} for the two-dimensional generalization. 
The idea is to reduce the eigenvalue problem in
eq.~\eqref{eq:schrodinger}, which depends on four independent radial
variables, to a set of coupled differential equations, which depend on
one radial and three angular variables. This leads to 
\begin{align}
    \label{eq:Schrodinger_HH}
    \begin{aligned}
        \Bigg\{ -\frac{1}{\rho}\left[ \frac{\partial^2}{\partial \xi^2}
  -\frac{3}{4\xi^2} - \frac{L\left(L+2\right)}{\xi^2}\right]-\hat{E} \Bigg\}
  \, u_{\{L\}}^{ } \left(\xi \right) +  \sum_{\{L'\}}
  \mathcal{V}_{\{L,L'\}}^{ }\left(\xi \right) 
  \,u_{\{L'\}}^{ }\left(\xi \right)=0 
  \,,
    \end{aligned}
\end{align}
where $\xi$, defined in eq.~\eqref{eq:def_xi},  
is the radial variable, 
and $L$ is a non-negative integer number. 

The matrix elements
$\mathcal{V}_{\{L,L'\}}^{ }$ are defined in eq.~\eqref{eq:PME_app}. 
They are computed in the basis provided by the eigenvectors of the
hyperangular momentum operator, defined in eq.~\eqref{eq:Hangular}.
Let us stress that those eigenvectors are known analytically, see
eq.~\eqref{eq:HH1}, and therefore the matrix
elements $\mathcal{V}_{\{L,L'\}}^{ }$ can be determined 
numerically with a
moderate computational effort. 
Since we are looking
for the ground state of the system, it is natural to restrict
ourselves to the case of zero total angular momentum
(cf.\ the discussion above eq.~\eqref{eq:even_L}). 
This 
reduces the calculation of the matrix elements 
to states associated with
even values of $L$ only.

For the numerical evaluation, 
we do need to restrict to a finite number of
eigenstates of the hyperangular momentum, i.e.\  truncate the basis
up to a maximal value $L_{\rm max}^{ }$. The choice of $L_{\rm max}^{ }$ is
critical and there is no general prescription for this choice,
which may be, in general, potential-dependent. 

%
\begin{figure}
    \centering
    \includegraphics[width=0.5\linewidth]{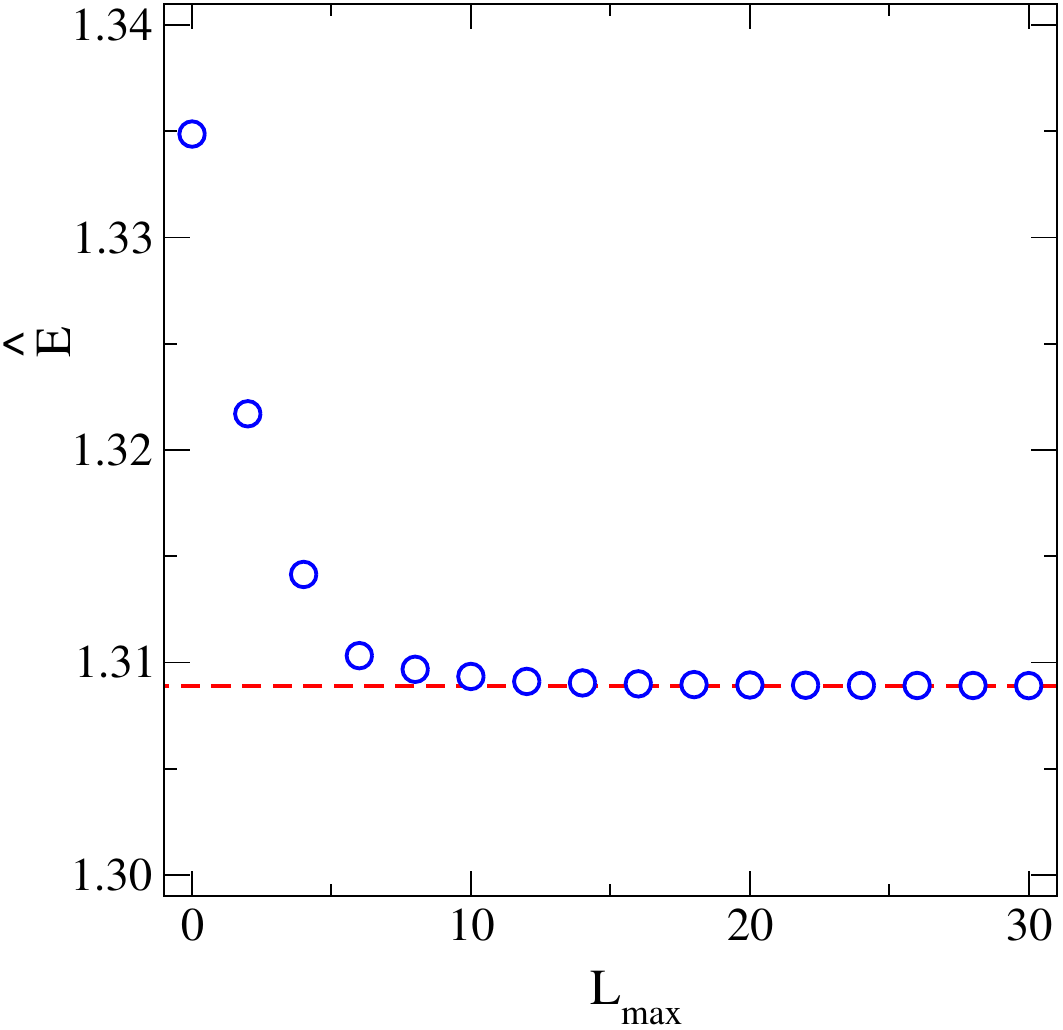}
    \caption{%
 Dependence of the ground-state
 eigenvalue~$\hat{E}$ (open blue circles) on the basis size
 $L_{\rm max}^{ }$. The calculation is restricted to even values 
 of $L$, in order to
 take into account eigenstates with zero total angular momentum. 
 The dashed red line represents the 
 infinite-$L^{ }_{\rm max}$ extrapolation.}
    \label{fig:GS-vs-L}
\end{figure}
%

In fig.~\ref{fig:GS-vs-L}, we show the ground-state
eigenvalue as a function of $L_{\text{max}}$. As expected, by
increasing the number of states, 
the result becomes increasingly stable. The relative
difference between the lowest eigenvalue extracted with 
$L_{\rm max}^{ }=28$ and with 
$L_{\rm max}^{ }=30$ is 
$\sim 10^{-6}_{ }$.
The final estimate for the ground-state eigenvalue 
for $N_\rmi{\sl f}^{ }=3$, is therefore
\begin{align}
\label{eq:eig}
    \hat{E} \, \overset{N_\rmii{\sl f}\,=\,3}{=} \, 1.309 \, .
\end{align}

As a further cross-check, we have
discretized the Hamiltonian on a mesh grid. The two relative 
positions from eq.~\eqref{eq:dimensionless} live in 
Cartesian dimensions of extent $N a$ each, where $a$ is a lattice
spacing (in units of $m_\rmii{E}^{-1}$). 
The wave functions are then $N^4_{ }$-component vectors, 
and the Hamiltonian is an $N^4_{ }\times N^4_{ }$-matrix. For the 
Laplacian a 5th order discretization is adopted. We extrapolate
separately $a\to 0$ with $N a$ fixed (continuum limit), 
and $N a \to \infty$ (infinite-volume limit), checking that 
the end result is independent of the ordering of these limits.
Including values in the ranges $N = 20...64$ 
and $a m^{ }_\rmii{E}=0.16...0.75$, 
we find that the results are consistent with an approximately
quadratic dependence on $a$. The continuum limit yields a result 
compatible with eq.~\eqref{eq:eig}, up to all digits shown.

Inserting eq.~\eqref{eq:eig} into eq.~\eqref{eq:Ehat_def}, 
leads to our NLO estimate of the baryonic screening masses, 
\begin{align}
\label{eq:E_1loop}
    \min\{ E \} 
    \; \overset{N_\rmii{\sl f}\,=\,3}{=} \; 
    3\pi T
    \,\bigg[\,
      1 + \bigg(\,
           \frac{1}{6\pi^2_{ }} + \frac{2\hat E}{9\pi^2_{ }}
          \,\biggr) g^2_{ }
        + O(g^3_{ })
    \,\bigg]
   \; \approx \;
   3\pi T \,  
   \bigl[\,  1+ 0.046 \, g^2_{ } + O(g^3_{ })
   \,\bigr] \, .
\end{align}
The NLO term represents 
a $\sim 4.6\%$ positive correction to the free-theory
result $3\pi T$ at the electroweak scale, 
i.e.\  $g^2_{ }\approx 1$~\cite{Laine:2005ai}. 

%
\section{Conclusions}
\label{sec:conclusions}

We have computed the NLO correction to a baryonic screening mass
in thermal QCD. The computation was carried out in the framework
of a dimensionally reduced effective theory, 
where the lowest Matsubara modes of massless quarks appear as
non-relativistic fields, 
interacting with zero Matsubara mode gluons. 

The NLO correction in eq.~\eqref{eq:E_1loop}
amounts to a $\sim 4.6 \%$ increase of 
the free-theory value $3\pi T$ at $g^2_{ }\approx 1$. 
For comparison, we note that an analogous
computation for the mesonic screening masses yields 
a $\sim 3\%$ positive
correction in the static sector~\cite{Laine:2003bd}, 
while in the non-static sector the effect is 
$\sim 5\%$~\cite{Brandt:2014uda}. The
reason for the difference lies in the form of the static
potential in the corresponding Schr\"odinger equation. 
The larger correction in the baryonic and non-static
mesonic sectors can be traced back to the appearance
of the potential $V^+_{ }$ in eq.~\eqref{eq:pot}, which is more
energetic than $V^-_{ }$ at short distances. 

The computation of our paper was motivated by, 
and in turn provides motivation for, a non-perturbative
determination of baryonic screening masses 
on the lattice~\cite{Rescigno:2023tkq,Giusti:2024ohu}. 
By comparing the two results, we can attempt to 
estimate the domain in which NLO computations can 
function as a quantitative tool for 
understanding the underlying physics. 
As discussed in more detail in ref.~\cite{Giusti:2024ohu}, 
the upshot is that, 
assuming that the coefficients of the higher-order corrections
are small, the $O(g^2_{ })$ term can account for
$\sim 90\%$ of the difference between the non-perturbative result
and the free-theory value $3\pi T$ down to temperatures of a few GeV.
This suggests that baryonic observables could be more perturbative
than most other quantities that have been studied so far. Of course, 
to consolidate this picture, it would be valuable to explicitly 
determine the coefficients of some higher-order corrections.

%
\section*{Acknowledgements}

D.L.\ thanks the Institute for Theoretical Physics of the 
University of Bern for hospitality during initial stages of this work. 
This work was (partially) supported by ICSC -- 
Centro Nazionale di Ricerca in High Performance Computing, 
Big Data and Quantum Computing, funded by European Union -- 
NextGenerationEU.
We thank CINECA for providing us with computing time 
on Marconi (CINECA-INFN, CINECA-Bicocca agreements).

%
\appendix

%
\section{Dimensionally reduced effective theory}
\label{app:eft}

%
\subsection{Bosonic sector}

For QCD in four dimensions we adopt the conventions in appendix~B of
ref.~\cite{DallaBrida:2020gux}. At finite temperature, the gauge field
can be written as
\begin{align}
 \label{eq:gauge3d}
 A_\mu^{ }(x_0^{ },x) = g_\rmii{E}^{ } \sum_{n}
 e^{iw_n x_0} A_{\mu,n}^{ } (x)
 \,, \quad 
 x = (\mathbf{r},x^{ }_3)
 \,, 
\end{align} 
where $A_{\mu,n}^{ }(x)$ are its Matsubara modes, 
with $w_n^{ }=2\pi T n$ and
$n\in \mathbb{Z}$. The prefactor~$g_\rmii{E}^{ }$ on the 
right-hand side of
eq.~(\ref{eq:gauge3d}) has been chosen so as to recover the
conventional perturbative normalization. 

The Matsubara modes effectively acquire 
a mass proportional to $|w_n^{ }|$, and
therefore the lightest degrees of freedom 
in the theory are the 
Matsubara zero modes $A_{\mu,0}^{ }(x)$.  The latter
are the building blocks of 
the dimensionally reduced effective theory. 
For simplicity, we denote the Matsubara 
zero modes by $A_{\mu}^{ }(x)$. Note that the three-dimensional
field has the mass dimension $[A_\mu^{ }]=1/2$.

Restricting to the set of Matsubara zero modes, 
the gauge transformations are taken 
constant in $x_0^{ }$. In other words, the effective
action has to respect three-dimensional gauge
invariance. If $\Omega(x)$ is a time-independent SU(3) element,
the gauge fields transform as
\begin{align}
    A^\Omega_i(x) & =  \Omega(x) A_i^{ }(x) \Omega^\dagger_{ }(x)
 +\frac{i}{g_\rmii{E}} \Omega(x) \partial_i^{ } \Omega^\dagger_{ }(x)\; , \\
    A^\Omega_0(x) & = \Omega(x) A_0^{ }(x) \Omega^\dagger_{ }(x) \, .
\end{align}
The $0$-component transforms in the
adjoint representation of the group, and gauge invariance
does not protect it from obtaining a mass term.

Given this field content, the effective action
in eq.~\eqref{eq:EQCD} is readily obtained.
A discussion of higher-dimensional
operators can be found in ref.~\cite{Laine:2016hma}.

%
\subsection{Fermionic sector}
\label{app:nrqcd}

In order to derive the effective action in eq.~\eqref{eq:NRQCD}, 
it is useful to employ a special representation of 
the (Euclidean) Dirac matrices, 
\begin{align}
    \gamma_0^{ } \, = \,
    \begin{pmatrix}
        0 & \id \\
        \id & 0
    \end{pmatrix} \, , \quad
    \gamma_1^{ } \, = \,
    \begin{pmatrix}
        \sigma_2^{ } & 0\\
        0 & -\sigma_2^{ }
    \end{pmatrix}\, , \quad
    \gamma_2^{ } \, = \,
    \begin{pmatrix}
        -\sigma_1^{ } & 0\\
        0 & \sigma_1^{ }
    \end{pmatrix}\, , \quad
    \gamma_3^{ } \, = \,
    \begin{pmatrix}
        0 & -i\id\\
        i\id & 0
    \end{pmatrix} \,,
    \label{eq:dirac}
\end{align}
where $\sigma_i^{ }$ are the $2\times 2$ Pauli matrices.
These imply
\begin{align}
    \gamma_5^{ } \, = \,
    \begin{pmatrix}
        -\sigma_3^{ } & 0\\
        0 & \sigma_3^{ }
    \end{pmatrix}\, , \quad
    C \, = i\gamma_0^{ }\gamma_2^{ } \, = \,
    \begin{pmatrix}
        0 & i\sigma_1^{ }\\
        -i\sigma_1^{ } & 0
    \end{pmatrix} \quad \text{and} \quad
    C\gamma_5^{ } \, = \,
    \begin{pmatrix}
        0 & \sigma_2^{ }\\
        \sigma_2^{ } & 0
    \end{pmatrix} \, ,
    \label{eq:clifford}
\end{align}
 and 
where $C$ is a hermitean charge conjugation.
With this representation the Dirac operator becomes
\begin{align}
    \gamma_0^{ } \slashed{D} \, = \,
    \begin{pmatrix}
        D_0^{ }+iD_3^{ } \;&\; -\epsilon_{kl}^{ } D_k^{ } \sigma_l^{ }\\
        \epsilon_{kl}^{ } D_k^{ } \sigma_l^{ } \;&\; D_0^{ }-iD_3^{ }
    \end{pmatrix} \, ,
    \label{eq:g0p}
\end{align}
where $k,l=1,2$ and $\epsilon_{kl}$ is the antisymmetric Levi-Civita
symbol with $\epsilon_{12}^{ }=1$. Similarly to eq.~\eqref{eq:gauge3d}, by
using the notation of ref.~\cite{Laine:2003bd}, we introduce the
three-dimensional two-components fields $\psi^{\uparrow}_{n}$ and
$\psi^{\downarrow}_{n}$, which are eigenstates
of $\gamma^{ }_0\gamma^{ }_3$. In the chosen basis,  
they are related to the usual
four-dimensional fermionic field $\psi$ by
\begin{align}
\label{eq:chiphi}
    \psi (x_0,x) =  \sqrt{T}\, \sum_{n} e^{i k_n x_0}
    \begin{pmatrix}
        \psi^{\uparrow}_{n} (x)\\
        \psi^{\downarrow}_{n} (x)
    \end{pmatrix} \, ,
\end{align}
where $n$ labels the fermionic Matsubara frequencies, 
$k_n^{ }=\pi T (2n+1)$ with $n\in \mathbb{Z}$. 
Inserting into the action of a single massless
fermion, $S_\psi^{ }$, we obtain\hspace*{0.3mm}\footnote{%
 For the notation we recall footnote~\ref{fn:bar}.
 }
\begin{align}
\begin{aligned}
\label{eq:dirac2}
 S_\psi^{ } \, = i \sum_n \int \! {\rm d}^3_{ } x \, 
 \Big\{ \bar{\psi}^{\uparrow}_{n} \left[k_n^{ } - g_\rmii{E}^{ } A_0^{ }
  + D_3^{ } \right]&\psi^{\uparrow}_{n} + \bar{\psi}^{\downarrow}_{n}
              \left[k_n^{ } 
  - g_\rmii{E}^{ } A_0^{ } -D_3^{ } \right]\psi^{\downarrow}_{n}\\
    &-i \bar{\psi}^{\downarrow}_{n} \epsilon_{kl}^{ }D_k^{ }\sigma_l^{ }
  \psi^{\uparrow}_{n}+i\bar{\psi}^{\uparrow}_{n}
  \epsilon_{kl}^{ } D_k^{ } \sigma_l^{ } \psi^{\downarrow}_{n}\Big\} 
  \,.
\end{aligned}
\end{align}
For a generic interpolating operator $O(y)$, 
the corresponding equations of motion are 
\begin{align}
\begin{aligned}
\label{eq:dirac3}
  i &\expval{ \Bigl\{\, 
  \left[k_n^{ } - g_\rmii{E}^{ } A_0^{ }
 + D_3^{ } \right]\psi^{\uparrow}_{n}
 + i\epsilon_{kl}^{ } D_k^{ } \sigma_l^{ } \psi^{\downarrow}_{n}
    \,\Bigr\} (x)\, O(y)}
   = \expval{\frac{\delta O(y)}{\delta
      \bar{\psi}^{\uparrow}_{n}(x)}}\, ,\\[0.25cm]
  i &\expval{ \Bigl\{\,
  \left[k_n^{ }  - g_\rmii{E}^{ } A_0^{ } - D_3^{ }\right]
 \psi^{\downarrow}_{n} 
 - i\epsilon_{kl}^{ }D_k^{ }\sigma_l^{ } \psi^{\uparrow}_{n}
    \,\Bigr\} (x)\, O(y)}
   = \expval{\frac{\delta O(y)}{\delta
      \bar{\psi}^{\downarrow}_{n}(x)}}\, ,
\end{aligned}
\end{align}
and analogously if we vary with respect to 
${\psi}^{\uparrow}_{n}$ and
${\psi}^{\downarrow}_{n}$.  

By inspecting the free limit, it is clear that 
for each Matsubara mode there is a mass term equal to
$k_n^{ }$. As a consequence, the lightest modes are in the sectors $n=0$
and $n=-1$. Considering propagation in the positive-$x^{ }_3$
direction and looking at the relative sign between the mass term and
the kinetic term, the light modes are $\psi^{\uparrow}_{0}$ and
$\psi^{\downarrow}_{-1}$. By solving the equations of motion for
$\psi^{\uparrow}_{0}$ and $\psi^{\downarrow}_{-1}$ at
$O(g_\rmii{E}^2)$, we obtain the equations of motion in
eqs.~(\ref{eq:EOM_chi}) and (\ref{eq:EOM_phi}), where we have dubbed
$\chi=\psi^{\uparrow}_{0}$ and $\phi=\psi^{\downarrow}_{-1}$,
respectively.

%
\begin{table}[t]
    \centering
    \begin{tabular}{c|c}
        field/operator & {power counting}\\[1mm]
        \hline\\[-3mm]
        $\chi,\phi$ & $m_\rmii{E}^{ }$\\[1mm]
        \hline\\[-3mm]
        $\partial_3^{ }\sim \Delta E$ & $g^2_\rmii{E}$ \\[1mm]
        \hline\\[-3mm]
        $\nabla_\mathbf{r}^{2} 
        = \nabla_\perp^2 \sim \mathbf{p}^2_{ }$ & $m_\rmii{E}^2$ \\[1mm]
        \hline\\[-3mm]
        ${\rm d}^3_{ }x$ & $g_\rmii{E}^{-2} m_\rmii{E}^{-2}$
    \end{tabular}
    \caption{Power counting for the fermionic fields in 
  eq.~\eqref{eq:NRQCD} and for their variation in longitudinal  
  ($x^{ }_3$) and transverse ($  \mathbf{r} = x^{ }_\perp$)
  directions.}
    \label{tab:power_counting}
\end{table}
%

In order to determine NLO corrections to hadronic correlation 
functions, it is helpful to establish the power
counting deriving from the three-dimensional 
action. The relevant scales of the system are  
$m_\rmii{E}^{ }$, $g_\rmii{E}^2$ and 
the thermal quark mass $k_0^{ }\approx M$, originating from 
the lowest fermionic Matsubara frequency. 
The power-counting rules are given in table
\ref{tab:power_counting}. As a consequence, the action in
eq.~\eqref{eq:NRQCD} is obtained by taking into account spinor fields
in the lowest Matsubara sector, 
including terms up to $O(g_\rmii{E}^0)$, 
and labelling, at variance with appendix~B of
ref.~\cite{DallaBrida:2020gux}, 
the $N_\rmi{\sl f}^{ }=3$ flavours as {\sl f}$\; =u,d,s$.

%
\section{Fermionic propagators in the effective theory}
\label{app:prop}

In coordinate space, by making explicit the source and the sink
locations, the propagator from eq.~\eqref{eq:free_prop1} becomes
\begin{align}
\label{eq:prop_free_coord}
    S^{(0)}_\chi(\mathbf{x},x_3^{ };\mathbf{y},y_3^{ }) =
   \frac{-i \theta(x^{ }_3 - y^{ }_3 ) T \id }{2 ( x^{ }_3-y^{ }_3 )}
  \exp \left[ - M ( x^{ }_3-y^{ }_3 )
  - \frac{ \pi T \left(\mathbf{x}-\mathbf{y} \right)^2}
    {2\left( x_3-y_3\right)} \right] \, ,
\end{align}
and similarly but with opposite sign for $\phi$. 
At the next order, from eq.~\eqref{eq:S_chi}, 
the propagator for $\chi$ reads
\begin{align}
\begin{aligned}
    S^{(1)}_{\chi} (\mathbf{x},x_3^{ }; \mathbf{y},y_3^{ })
    \; = \;
    i \int \! {\rm d} z^{ }_3 \! 
    \int \!{\rm d}^2_{ } \mathbf{z} \, 
    S^{(0)}_\chi&(\mathbf{x},x_3^{ };\mathbf{z},z^{ }_3)
    \left[ iA_3^{ }+A_0^{ } \right](\mathbf{z},z^{ }_3)
    S^{(0)}_\chi(\mathbf{z},z^{ }_3;\mathbf{y},y_3^{ })  \, ,
    \label{eq:S_chi_1}
\end{aligned}
\end{align}
implying that $\chi$ propagates freely
from the source to the position $(\mathbf{z},z^{ }_3)$, 
where it emits a gluon, 
and then again freely propagates up to the sink position. 
By inserting eq.~\eqref{eq:prop_free_coord} into 
eq.~\eqref{eq:S_chi_1} and taking $(\mathbf{y},y_3^{ })=(\mathbf{0},0)$,
the result can be rewritten as
\begin{align}
    \begin{aligned}
        S^{(1)}_\chi(\mathbf{x},x_3^{ };\mathbf{0},0)
  \;&= \; S^{(0)}_\chi(\mathbf{x},x_3^{ };\mathbf{0},0)
  \int_0^{x_3} \! {\rm d} z^{ }_3\, \int \! {\rm d}^2_{ } \mathbf{z}
  \, \frac{x^{ }_3 T}{2 (x^{ }_3-z^{ }_3)z^{ }_3} \\[0.125cm]
  & \times\exp\left[ -\frac{1}{2}
   \frac{\pi T}{\left(x^{ }_3-z^{ }_3 \right)}\frac{x^{ }_3}{z^{ }_3}
  \left( \mathbf{z}-\frac{z^{ }_3}{x^{ }_3}\mathbf{x} \right)^2\right]
  \left[ i A_3^{ }+A_0^{ } \right] (\mathbf{z},z^{ }_3) \, .
    \end{aligned}
\end{align}
{}From table~\ref{tab:power_counting} we see that the prefactor
of the quadratic dependence is parametrically of order 
$
 \pi T x^{ }_3 / [(x^{ }_3 - z^{ }_3) z^{ }_3] \sim \pi T g_\rmii{E}^2
$, 
implying that fermions probe the variation of gauge fields 
at distances 
$
 |\mathbf{z} - z^{ }_3 \mathbf{x} / x^{ }_3 |
 \sim 1/(g T) \sim 1/ m^{ }_\rmii{E}
$.
For the static potential, we need the $x^{ }_3\to \infty$ limit, 
and then only the position of the gauge field at 
$z^{ }_3\sim x^{ }_3$ matters. Then the prefactor is large, 
and we may evaluate the integral in the saddle point approximation, 
yielding
\begin{align}
    \begin{aligned}
        S^{(1)}_\chi(\mathbf{x},x_3^{ };\mathbf{0},0)
 \; \simeq \; 
 S^{(0)}_\chi(\mathbf{x},x_3^{ };\mathbf{0},0)  \,
 \int_0^{x_3} \! {\rm d}z^{ }_3\, \left[ iA_3^{ }+A_0^{ }\right]
 \left(\frac{z^{ }_3}{x^{ }_3}\mathbf{x} ,z^{ }_3\right)
 \, ,
    \end{aligned}
\end{align}
justifying eq.~\eqref{eq:ge_prop1}.

%
\section{Evaluation of the static potential}
\label{app:int}

We show here details for the manipulation of eq.~\eqref{eq:EOMge}.
The free-theory gauge field propagator reads
\begin{align}
 \label{eq:gauge_prop_mu}
 \bigl\langle\, A^a_\mu(x) A^b_\nu(0) \,\bigr\rangle^{ }_0 
 = \delta^{ab}_{ } \Delta_{\mu\nu}^{ }(x)
 \,, 
\end{align}
where $\mu,\nu=0,\dots,3$, 
$a,b=1,\dots, 8$ and, in Feynman gauge,  
\be
    \Delta_{\mu\nu}^{ }(x)  \; = \;  
    \int\! \frac{{\rm d}^3_{ } p}{(2\pi)^3_{ }} \, 
    e^{i p \cdot x} \, \biggl[\, 
    \frac{\delta^{ }_{\mu 0}\delta^{ }_{\nu 0}}{p^2+m_\rmii{E}^2}
    \, + \, 
    \frac{\delta^{ }_{\mu i}\delta^{ }_{\nu i}}{p^2+\lambda^2_{ }} 
    \, \biggr] 
    \;, \quad p=(\mathbf{p},p^{ }_3)
    \,.
\label{eq:gauge_prop_i}
\ee
The Debye mass $m_\rmii{E}^{ }$ is given in eq.~\eqref{eq:ge}, and 
$\lambda$ is an auxiliary mass parameter, used as an IR regulator.  
The Wick
contractions following from eq.~\eqref{eq:EOMge}
lead to the colour algebra 
\be
 \epsilon^{abc}_{ }\epsilon^{gfe}_{ }
 (T^d_{ }T^d_{ })^{ag}_{ }\delta^{bf}_{ }\delta^{ce}_{ } 
 = 8 
 \;, \quad
 \epsilon^{abc}_{ }\epsilon^{gfe}_{ }
 (T^d_{ })^{ag}_{ }(T^d_{ })^{bf}_{ } \delta^{ce}_{ } 
 = -4 
 \,, 
\ee
where $T^{d}_{ }$ are Hermitean generators of SU(3), 
normalized as $\tr (T^c_{ }T^d_{ }) = \delta^{cd}_{}/2$. 
Subsequently, we are left over with the time-integrals
\begin{align}
 {\cal V}^{\pm}_{ }(\textbf{r}_1^{ },\textbf{r}_2^{ },x_3^{ })\equiv
  -\int_0^{x_3} \! {\rm  d}z^{ }_3\, \Big[\Delta_{33}^{ }
 \Bigl(\, \textbf{r}_1^{ }-\textbf{r}_2^{ }
 +\frac{z^{ }_3}{x^{ }_3}\textbf{r}_2^{ } ,z^{ }_3
 \,\Bigr) 
 \mp
 \Delta_{00}^{ }
 \Bigl(\, \textbf{r}_1^{ }-\textbf{r}_2^{ }
 +\frac{z^{ }_3}{x^{ }_3}\textbf{r}_2^{ },z^{ }_3
 \,\Bigr)\Big]
 \,. \label{eq:calVpm}
\end{align}
The propagators are written in a Fourier representation, 
like in eq.~\eqref{eq:gauge_prop_i}.
In the limit $x_3^{ }\rightarrow\infty$, 
noting that 
\be
 \lim_{\delta\to 0^+_{ } }
 \lim_{x^{ }_3\to\infty}
 \int_0^{x^{ }_3} \! {\rm d}z^{ }_3 \, 
 e_{ }^{\bigl[ i (\frac{\textbf{p}\cdot\textbf{r}}{x_3} + p^{ }_3)
 - \delta \bigr] z^{ }_3}
 \; = \; 
 \lim_{\delta\to 0^+_{ } }
 \lim_{x^{ }_3\to\infty}
 \frac{i}{ \frac{\textbf{p}\cdot\textbf{r}}{x_3} + p^{ }_3 + i \delta }
 \; = \; \mathbbm{P}\biggl( \frac{i}{p^{ }_3} \biggr)
   + \pi \delta(p^{ }_3)
 \;, 
\ee
where $\mathbbm{P}$ denotes a principal value, 
and pulling the time integral inside the Fourier representation, 
the Fourier transform becomes
\be
 \int_{\textbf{p}} \int_{-\infty}^{\infty} \! \frac{{\rm d}p^{ }_3}{2\pi}
 \int_0^{x^{ }_3} \! {\rm d}z^{ }_3 \, 
 \frac{
 e_{ }^{i \textbf{p}\cdot
       (\textbf{r}^{ }_1 - \textbf{r}^{ }_2 + \frac{z_3}{x_3}\textbf{r}_2)
       + i p^{ }_3 z^{ }_3}
      }{f(\textbf{p}^2_{ },p_3^2)}
 \stackrel{x^{ }_3\to\infty}{\longrightarrow}
 \frac{1}{2} 
 \int_{\textbf{p}} 
 \frac{e_{ }^{i\textbf{p}\cdot(\textbf{r}^{ }_1 - \textbf{r}^{ }_2)}}
 {f(\textbf{p}^2_{ },0)}
 \;. \label{eq:large_x_3}
\ee
Carrying out the two-dimensional momentum integral, 
the potential reads
\begin{align}
\lim_{x_3\rightarrow\infty}
    {\cal V}^{\pm}_{ }(\textbf{r}_1^{ },\textbf{r}_2^{ },x_3^{ }) =
    \frac{1}{4\pi} 
 \left[\ln\left(\frac{\lambda\, r_{12}}{2}\right)
 +  \gamma_\rmii{E}^{ } \pm K_0^{ }(m_\rmii{E}^{ } r_{12}^{ })\right]
 \,,
\end{align}
where $r_{12}^{ }\equiv |\textbf{r}_1^{ }-\textbf{r}_2^{ }|$,
$\gamma_\rmii{E}^{ }$ is the Euler-Mascheroni constant, 
and $K_0^{ }$ a modified Bessel function.
The combinations that appear in eq.~\eqref{eq:Kstor} are defined as 
\begin{align}
 V^\pm_{ } (r_{12}^{ }) \;\equiv\;  \frac{4 g^2_\rmii{E} }{3} 
 \lim_{\lambda\rightarrow 0} \lim_{x_3^{ }\rightarrow\infty}
 \bigl[
    {\cal V}^{\pm}_{ }(\textbf{r}_1^{ },\textbf{r}_2^{ },x_3^{ })
  + {\cal V}^{\pm}_{ }(\textbf{r}_2^{ },\textbf{r}_1^{ },x_3^{ })
  - {\cal V}^{+}_{ }(\textbf{r}_1^{ },\textbf{r}_1^{ },x_3^{ })
  - {\cal V}^{+}_{ }(\textbf{r}_2^{ },\textbf{r}_2^{ },x_3^{ })
  \bigr]\,.
\end{align}
By using the expansion of $K_0^{ }(x)$ for small $x$, 
it is straightforward to show that 
\be
 \lim_{x_3\rightarrow\infty} {\cal
 V}^{+}_{ }(\textbf{r}_1^{ },\textbf{r}_1^{ },x_3^{ }) = \frac{1}{4\pi}
 \ln\left(\frac{\lambda}{m_\rmii{E}}\right)
 \,. 
\ee
This then leads to the explicit expressions for $V^\pm_{ } (r)$
reported in eq.~(\ref{eq:Vpm}).

%
\section{Two-dimensional hyperspherical harmonics}
\label{app:HH}

Let us consider a system of three particles of mass~$m$
which are constrained to move in a two-dimensional plane. 
We assume those particles to interact through mutual
two-body potentials which depend only on relative separations.
The Schr\"odinger equation for such a system is
\begin{align}
    \left[ -\frac{1}{2m} \sum_{i=1}^3 \nabla_{\mathbf{r}_{i}^{ }}^2
   + \sum_{j>i}^{3} V(r_{ij}^{ }) \right] 
  \psi(\mathbf{r}_1^{ },\mathbf{r}^{ }_2,\mathbf{r}^{ }_3)
  \, = \, E\, \psi (\mathbf{r}_1^{ },\mathbf{r}^{ }_2,\mathbf{r}^{ }_3) \, ,
\end{align}
where $r_{ij}^{ } \equiv |\mathbf{r}_i^{ }-\mathbf{r}_j^{ }|$. 
By performing the
change of variables in eq.~\eqref{eq:Jacobi}, the Laplace operator
can be written as
\begin{align}
    \frac{1}{2}\Bigl [\nabla_{{\mathbf{r}}_1}^2
    +\nabla_{{\mathbf{r}}_2}^2
    +\nabla_{{\mathbf{r}}_3}^2 \Bigr ] 
    \, = \, \frac{\nabla_{{\mathbf{R}}}^2}{6}
    +\nabla_{{\bsxi}_1}^2+\nabla_{{\bsxi}_2}^2 
   \, , \label{eq:laplacian}
\end{align}
where ${\mathbf{R}}$ is the center-of-mass coordinate. 
In such a way, the kinetic energy 
can be separated into center-of-mass and relative motions. 
The Schr\"odinger equation for the relative motion
can be written in terms of the new coordinates as
\begin{align}
\label{eq:sch_HH1}
    \lb - \frac{1}{m}\left( \nabla_{{\bsxi_1}}^2
    +\nabla_{{\bsxi}_2}^2 \right)
    +{V}\left( {\bsxi_1^{ }},{\bsxi_2^{ }}\right)
   \rb {\Psi}\left({\bsxi_1^{ }},{\bsxi_2^{ }}\right)
  \, = \, E\, \Psi\left({\bsxi_1^{ }},{\bsxi_2^{ }}\right) \,.
\end{align}
By going to polar coordinates, i.e.\ 
\begin{align}
\label{eq:azimuth}
\begin{cases}
    {\xi}_1^{(1)} \, = \, {\xi}_1^{ } \cos \theta_1^{ } \\
    {\xi}_1^{(2)} \, = \, {\xi}_1^{ } \sin \theta_1^{ }
\end{cases}
\qquad
\begin{cases}
    {\xi}_2^{(1)} \, = \, {\xi}_2^{ } \cos \theta_2^{ } \\
    {\xi}_2^{(2)} \, = \, {\xi}_2^{ } \sin \theta_2^{ }
\end{cases}
\, , 
\end{align}
where ${\xi}_i^{ }$ refers to the absolute value of the 2-vector
${\bsxi}_i^{ }$ and the superscript $(j)$ refers to the components
in the two-dimensional plane, the Laplace operator 
can be expressed as
\begin{align}
\label{eq:lapl_polar}
    \nabla^2_{{\bsxi}_1}+\nabla^2_{{\bsxi}_2} 
    \, = \, \frac{\partial^2}{\partial {\xi}_1^2} 
     +\frac{1}{{\xi}_1} \frac{\partial}{\partial {\xi}_1} 
     +\frac{1}{{\xi}_1^2} \frac{\partial^2}{\partial \theta_1^2} 
    \, + \, \frac{\partial^2}{\partial {\xi}_2^2} 
     +\frac{1}{{\xi}_2} \frac{\partial}{\partial {\xi}_2} 
     +\frac{1}{{\xi}_2^2} \frac{\partial^2}{\partial \theta_2^2} \, .
\end{align}
Both for ${\xi}_1^{ }$ and ${\xi}_2^{ }$ the angular differential operator 
corresponds to the square of the two-dimensional 
quantum-mechanical angular momentum operator, 
\begin{align}
   \mathbf{L}_i^{ } \, \equiv \, (-i) \frac{\partial}{\partial \theta_i} 
   \qquad \textrm{for} \quad i=1,2 \, .
\end{align}
We also note that 
the scalar products following from 
eqs.~\eqref{eq:Jacobi} and ~\eqref{eq:xis} read 
\be
 r^{ }_{12} = 
 \frac{\sqrt{\xi_1^2 + 3 \xi_2^2
   + 2 \sqrt{3} {\bsxi}_1^{ }\cdot {\bsxi}^{ }_2
                        }}{2}
 \;, \quad
 r^{ }_{13} = \xi_1^{ }
 \;, \quad
 r^{ }_{23} = 
 \frac{\sqrt{\xi_1^2 + 3 \xi_2^2 
   - 2 \sqrt{3} {\bsxi}_1^{ }\cdot {\bsxi}^{ }_2
                        }}{2}
 \,,
 \label{eq:distances_xi}
\ee
where
$
 {\bsxi}_1^{ }\cdot {\bsxi}^{ }_2
 = 
 \xi^{ }_1 \xi^{ }_2 \cos(\theta^{ }_1 - \theta^{ }_2)
$.

So far, our coordinates comprise two radial extension ${\xi}_i^{ }$
and two angles $\theta_i^{ }$. The idea underlying the 
hyperspherical
harmonics
method is to write the system, by a suitable change of variables, in
terms of one radial and three angular variables. This is achieved by
going to the so-called \textit{hyperspherical coordinates}, 
\begin{align}
    \begin{cases}
        {\xi}_1^{ } \, = \, \xi \sin \phi\\
        {\xi}_2^{ } \, = \, \xi \cos \phi
    \end{cases} \, , \label{eq:def_xi}
\end{align}
where $\xi$ is called the \textit{hyper-radius} and the angular variable
$\phi$ describes the relative length between the two vectors
${\bsxi}_1^{ }$ and ${\bsxi}_2^{ }$, i.e.\  $\phi \in [0,\pi/2]$. 
Together with
$\theta_1^{ }$ and $\theta_2^{ }$, $\phi$ forms a set of so-called
\textit{hyper-angles} 
$\Omega_4^{ } \equiv \lb \theta_1^{ }, \theta_2^{ },\phi \rb$. 
With this change of variables, the Laplace operator in
eq.~\eqref{eq:lapl_polar} becomes
\begin{align}
\label{eq:laplacian_HH}
   \nabla^2_{\xi}\, \equiv \,\nabla^2_{{\bsxi}_1}
  +\nabla^2_{{\bsxi}_2} \, = \, \frac{\partial^2}{\partial \xi ^2} 
  + \frac{3}{\xi} \frac{\partial}{\partial\xi}
  - \frac{\mathcal{L}^2\left(\Omega_4\right)}{\xi^2} \, ,
\end{align}
where we introduced the hyperangular momentum operator
\begin{align}
\label{eq:Hangular}
    \mathcal{L}^2_{ }\left( \Omega_4^{ } \right) 
 \; \equiv \; -\left[ \frac{\partial^2}{\partial^2 \phi} 
 +2\cot{2\phi}\, \frac{\partial}{\partial \phi} 
 -  \frac{\mathbf{L}^2_1}{\sin^2 \phi} 
 -  \frac{\mathbf{L}^2_2}{\cos^2 \phi} \right] \, .
\end{align}

In complete analogy with the angular momentum operator, we can define
the eigenvalues and eigenstates of the hyperangular momentum
operator by assuming that $\xi^L_{ } \mathcal{Y}\left(\Omega_4^{ }\right)$ is
a harmonic function. This leads to
\begin{align}
\label{eig_H}
  \mathcal{L}^2_{ }\left( \Omega_4^{ } \right) 
  \mathcal{Y}\left( \Omega_4^{ } \right) \, = \, 
 L\left( L+2 \right)\mathcal{Y}\left( \Omega_4^{ } \right) \, ,
\end{align}
where $L$, in analogy with the angular momentum quantum number, is
called the \textit{grand orbital} quantum number. Given the form of the
operator in eq.~\eqref{eq:Hangular}, the corresponding eigenvectors
can be written as
\begin{align}
    \label{eq:HH1}
     \mathcal{Y}_{L}^{ }\left( \Omega_4^{ } \right) 
   \, = \, \mathcal{P}_L^{l_1 l_2} \left( \phi \right) Y_{l_1}^{ }
  \left( \theta_1^{ }\right) Y_{l_2}^{ } \left( \theta_2^{ }\right) \, ,
\end{align}
where $Y_{l_i}^{ }(\theta)$ are the eigenvectors 
of the $\mathbf{L}_i^{ }$ operators, i.e.\ 
\begin{align}
     Y_{l_i}^{ } \left( \theta_i^{ } \right) \, = \, 
  \frac{e^{il_i\theta_i}}{\sqrt{2\pi}} \qquad \text{with}
  \qquad l_i^{ }\, \in \, \mathbb{Z} \, .
\end{align}
By looking for a solution of the type
\begin{align}
\label{eq:phi_eig}
    \mathcal{P}_L^{l_1 l_2} \left( \phi \right) 
    \, = \, 
    \mathcal{N}^{l_1 l_2}_{L}
    \left( \sin \phi \right)^{|l_1|}_{ } 
    \left( \cos \phi \right)^{|l_2|}_{ } 
   g\left( \cos 2\phi \right) 
   \,, 
\end{align}
and by performing the change of variables $x=\cos 2\phi$, 
the differential equation in eq.~\eqref{eig_H} can be written as
\begin{align}
    \left(1-x^2\right) \frac{\partial^2 g}{\partial x^2}
     \, + \, \big[ |l_2^{ }|-|l_1^{ }|&-(|l_1^{ }|+|l_2^{ }|+2)\,x\,
     \big]\frac{\partial g}{\partial x}
    + \, \left( n+|l_1^{ }|+|l_2^{ }|+1 \right) n \, g 
 \, = \, 0 \,, \label{eq:g}
\end{align}
where $n$ is related to the quantum numbers $l_1^{ }, l_2^{ }$ 
and $L$ by 
\begin{align}
\label{eq:L}
    L \, = \, 2n +|l_1^{ }| + |l_2^{ }|
   \qquad \text{with}\qquad n \in \mathbb{N} \, .
\end{align}
Equation~\eqref{eq:g} has analytical solutions in terms of Jacobi
polynomials, 
\begin{align}
   g(x) \, = \, P_n^{l_1 l_2} (x) 
   \, = \, \frac{\left( 1+|l_1| \right)_n}{n!} 
   \, \leftidx{_2}{F}{_1} \left( -n, n+|l_1^{ }|+|l_2^{ }|+1; |l_1^{ }|+1; 
 \frac{1-x}{2} \right) \, ,
\end{align}
where $(\cdot)_n^{ }$ is the Pochhammer symbol and $\leftidx{_2}{F}{_1}$
is the hypergeometric function. These polynomials are orthogonal in
the interval $[-1,1]$ with respect to the weight function
$\left(1-x\right)^{|l_1|} \left(1+x\right)^{|l_2|}$, 
and their normalization constant is given by
\begin{align}
  \int_{-1}^{1} \! {\rm d}x \, P_n^{l_1 l_2}(x)  
  P_{n'}^{l_1 l_2}(x) &\left(1-x\right)^{|l_1|} 
  \left(1+x\right)^{|l_2|}  \\
    &= \, \frac{\delta_{n,n'} \, 2^{|l_1| +|l_2| +1}}{2n +|l_1|+|l_2|+1}
  \frac{\Gamma\left( n+|l_1|+1 \right)
  \Gamma\left( n+|l_2|+1 \right)}{\Gamma\left(n+1\right)
  \Gamma\left( n+|l_1|+|l_2|+1 \right)} \;, \nonumber
\end{align}
where $\Gamma$ is the Euler gamma function. 

By inserting the solution into eq.~\eqref{eq:phi_eig}, 
and taking into account the volume element for the angular variables, 
$
 \diff
 \Omega_4^{ } \, \equiv \, \sin \phi \cos \phi 
 \, \diff \phi
 \, \diff \theta_1^{ } 
 \, \diff \theta_2^{ } 
$, 
the normalization of the eigenvectors in
eq.~\eqref{eq:phi_eig} is obtained by integrating over the hyperangle
$\phi$,
\begin{align}
  \bigl( \mathcal{N}_L^{l_1 l_2} \bigr)^2_{ } \int_0 ^{\frac{\pi}{2}}
 \! {\rm d}\phi \,
  \left( \sin \phi \right)^{2|l_1|+1} 
  \left( \cos \phi \right)^{2|l_2|+1}
  P_n^{l_1 l_2}(\cos 2\phi) 
  P_{n'}^{l_1 l_2}(\cos 2\phi) \, = \, \delta_{n,n'}^{ } \, .
\end{align}
This determines the normalization constant,
\begin{align}
    \mathcal{N}_L^{l_1 l_2} 
   \, = \, \left[ \frac{2\left(L+1\right) 
  \Gamma \left( n+1 \right) \Gamma \left(L -n+1\right)}
 {\Gamma\left(n+|l_1| + 1\right) 
 \Gamma\left( n + |l_2| +1\right)} \right]^{\frac{1}{2}}_{ } \, .
\end{align}
As a consequence, the hyperspherical harmonics ${\cal Y}_L^{ }$ 
satisfy the orthonormality relation
\begin{align}
\label{eq:orth}
   \int \! {\rm d} \Omega_4^{ } \,  \mathcal{Y}_L^* 
   \left( \Omega_4^{ } \right) \mathcal{Y}_{L'}^{ } 
   \left( \Omega_4^{ } \right) \, = \, \delta_{l_1^{ }, l_1'}^{ } 
   \delta_{l_2^{ }, l_2'}^{ }  \delta_{n,n'}^{ } \, .
\end{align}

The hyperspherical harmonics form a complete set 
on the surface described by the angles $\lb \phi, \theta_1^{ },
\theta_2^{ } \rb$ at fixed $\xi$. As a consequence, the wave functions
appearing in eq.~\eqref{eq:sch_HH1} can be expanded as
\begin{align}
\label{eq:exp}
   \Psi\left({\bsxi_1^{ }},{\bsxi_2^{ }}\right) 
   \, = \, \sum_{\{L\}} \Phi_{\{L\}}^{ }
  \left(\xi \right) \mathcal{Y}_{\{L\}}^{ }\left(\Omega_4^{ }\right) \, ,
\end{align}
where $\{L\}$ refers to any particular combination of quantum numbers
which leads to $L=2n+|l_1^{ }| +|l_2^{ }|$, 
and the expansion coefficients
$\Phi_{\{L\}}^{ }$ encode the hyper-radial dependence of the wave
function. We can get rid of the first derivative appearing
in eq.~\eqref{eq:laplacian_HH} by writing the expansion coefficient as
$\Phi^{ }_{\{L\}}\left(\xi \right) = 
 u^{ }_{\{L\}}\left( \xi \right)/\xi^{3/2}_{ }$. Therefore, by
using eq.~\eqref{eig_H}, the Schr\"odinger equation can be written as
\begin{align}
    \sum_{\{L'\}} \left\lbrace
   -\frac{1}{m}\left[ \frac{\partial^2}{\partial \xi^2} 
   -\frac{3}{4\xi^2} - \frac{L'\left(L'+2\right)}{\xi^2}\right]  
   + V\left(\xi,\Omega_4^{ }\right)-E \right\rbrace u_{\{L'\}}^{ }
    \left(\xi \right) \mathcal{Y}_{\{L'\}}^{ }
    \left(\Omega_4^{ }\right)\, = \, 0\, .
\end{align}
By multiplying on the left by
$\mathcal{Y}^*_{\{L\}}\left(\Omega_4^{ }\right)$ and integrating over 
$\Omega_4^{ }$, the orthogonality relation for the
hyperspherical harmonics, eq.~\eqref{eq:orth}, 
leads to the final form
\begin{align}
    \left\lbrace -\frac{1}{m}\left[ \frac{\partial^2}{\partial \xi^2}
   -\frac{3}{4\xi^2} - \frac{L\left(L+2\right)}{\xi^2}\right] 
  \, - \, E \right\rbrace u_{\{L\}}^{ } \left(\xi \right) 
  \, + \, \sum_{\{L'\}} \mathcal{V}_{\{L,L'\}}^{ }\left(\xi \right)
  \,u_{\{L'\}}^{ }\left(\xi \right) =0
  \,,
\end{align}
where
\begin{align}
\label{eq:PME_app}
  \mathcal{V}_{\{L,L'\}}^{ }\left(\xi \right)
  \, = \, \int \! {\rm d}\,\Omega_4^{ }
  \,\mathcal{Y}_{\{L\}}^*\left(\Omega_4^{ }\right)
  {V}\left(\xi,\Omega_4^{ }\right) \mathcal{Y}_{\{L'\}}^{ } (\Omega_4^{ })
\end{align}
are the matrix elements computed in the basis of the
hyperspherical harmonics. 

In this way, the Schr\"odinger
equation has been reduced to a set of coupled one-dimensional
differential equations which are easier to solve numerically. With
this procedure, the bulk of the calculation is given by the computation
of the matrix elements by properly truncating the basis up
to a certain value $L_{\text{max}}$. The number of states to be
included can be reduced
by scrutinizing the physics of the problem of interest. 
Given that our potential $V$ only depends on 
$\theta^{ }_1 - \theta^{ }_2$
(cf.\ eq.~\eqref{eq:distances_xi}), 
the total angular momentum in the transverse plane, 
with its eigenvalue $l^{ }_1 + l^{ }_2$, commutes with the 
Hamiltonian and represents a conserved quantity. 
In order to extract the energy of the ground state, it is then 
sufficient to 
consider the smallest value of the total angular momentum, 
i.e.\ set  $ l_2^{ }=-l_1^{ } $ in eq.~\eqref{eq:L}.
This implies that the grand orbital quantum
number is restricted to even values only, 
\begin{align}
    L=2\left( n+|l_1^{ }| \right) \, .
    \label{eq:even_L}
\end{align}
Furthermore, given that the potential is invariant under
$\theta^{ }_1 \to \theta^{ }_1 + \pi$, 
we can restrict to wave functions symmetric under this ``parity'', 
and choose $l^{ }_1$ even. 
In general, any fixed~$L$ describes a set $\{ L\}$ 
of various $n$ and $l^{ }_1$.

\bibliographystyle{JHEP}
\bibliography{biblio04}

\end{document}